\documentclass[superscriptaddress,showpacs,aps]{revtex4}

\usepackage{graphicx}

\newcommand\e{\epsilon}

\renewcommand\l{\lambda}
\newcommand\m{\mu}

\newcommand\p{\pi}

\renewcommand\t{\tau}

\newcommand\f{\phi}

\newcommand\D{\Delta}

\newcommand{\lan}{\langle}
\newcommand{\ran}{\rangle}

\newcommand{\non}{\nonumber\\}

\newcommand{\be}{\begin{equation}}
\newcommand{\ee}{\end{equation}}
\newcommand{\bea}{\begin{eqnarray}}
\newcommand{\eea}{\end{eqnarray}}
\newcommand{\ba}[1]{\begin{array}{#1}}
\newcommand{\ea}{\end{array}}

\begin{document}

\title{The non-Abelian feature of parton energy loss 
in energy dependence of jet quenching in high-energy heavy-ion collisions}

\author{Qun Wang}
\affiliation{Department of Modern Physics, 
University of Science and Technology of China,
Hefei, Anhui 230026, People's Republic of China}
\affiliation{Institut f\"ur Theoretische Physik, 
Johann Wolfgang Goethe-Universit\"at,
Postfach 111932, 60054 Frankfurt am Main, Germany}

\author{Xin-Nian Wang}
\affiliation{Nuclear Science Division, MS70R0319,
Lawrence Berkeley National Laboratory, Berkeley, CA 94720}

\begin{abstract}
One of the non-Abelian features of parton energy loss is the
ratio $\Delta E_g/\Delta E_q=9/4$ between gluon and quark jets.
Since jet production rate is dominated by quark jets at high 
$x_T=2p_T/\sqrt{s}$ and by gluon jets at low $x_T$, high $p_T$
hadron suppression in high-energy heavy-ion collisions should 
reflect such a non-Abelian feature. Within a leading order perturbative 
QCD parton model that incorporates transverse expansion and Woods-Saxon 
nuclear distribution, the energy dependence of large $p_T\sim 5-20$ GeV/$c$
hadron suppression is found to be sensitive to the non-Abelian
feasture of parton energy loss and could be tested by data from
low energy runs at RHIC or data from LHC.

\end{abstract}

\pacs{12.38.Mh, 24.85.+p; 13.60.-r,25.75.-q}

\maketitle


One of the ultimate goal of the Relativistic Heavy-Ion Collider (RHIC) 
at Brookhaven National Laboratory is to produce the quark-gluon
plasma (QGP) by smashing two gold nuclei at the speed of light. 
The discovery of jet quenching effect 
\cite{Adcox:2001jp,Adler:2002xw,Adler:2002tq,Adler:2002ct} 
in central Au+Au collisions together with the observation 
of parton recombination 
\cite{Adler:2001bp,Adcox:2001mf,Adler:2002pba,Adcox:2002au,
Hwa:2002tu,Fries:2003vb,Greco:2003xt,Voloshin:2002wa}
and the early thermalization of the dense matter 
\cite{Adams:2003zg,Adams:2003am,Adler:2002pu,
Adler:2003kt,Stocker:1979bi,Csernai:1981nq,Ollitrault:1992bk,
Molnar:2001ux,Zhang:1999rs,Xu:2004mz} 
has provided clear evidence for the formation 
of strongly interacting partonic matter \cite{Gyulassy:2004zy,Jacobs:2004qv}. 
The observed jet quenching effect manifests
itself in several aspects of high $p_T$ hadron spectra, 
which include suppression of inclusive spectra in Au+Au
relative to $pp$ collisions \cite{Adcox:2001jp,Adler:2002xw}, 
disappearance of back-to-back correlation \cite{Adler:2002tq}
and the azimuthal anisotropy in non-central Au+Au collisions 
\cite{Adler:2002ct}. The absence of these jet quenching phenomena in $d+Au$
collisions \cite{Adams:2003im,Adler:2003ii,Back:2003ns,Arsene:2003yk} 
shows that they are due to final state interactions with the
produced strongly interacting matter.
Detailed analyses indicate that parton 
energy loss is the source for the observed jet quenching
\cite{Wang:2003aw,Wang:xy,Pluemer:dp,Wang:2000fq,
Gyulassy:2000gk,Eskola:2004cr}. 
The initial gluon density, which the parton energy loss
is proportional to, has been extracted from RHIC data
of central Au+Au collisions at $\sqrt{s}=200$ AGeV and
is about 30 times higher than that in a cold nucleus 
\cite{Wang:2003mm,Vitev:2004bh}.

The radiative parton energy loss incoporated in previous studies 
within a leading order (LO) perturbative QCD (pQCD) model 
\cite{Wang:1998ww,Wang:1998bh,Wang:2000fq,
Wang:2004yv,Gyulassy:2000gk,Vitev:2002pf}
has two basic non-Abelian features. One of them is the quadratic 
dependence on the total distance traversed by the propagating 
parton due to the non-Abelian Landau-Pomeranchuk-Midal (LPM) 
interference effect in gluon bremsstrahlung induced by mulitple 
scatterings in a {\it static} medium
\cite{Gyulassy:1993hr,Wang:1994fx,
Gyulassy:2000fs,Gyulassy:1999zd,Baier:2001yt,
Wiedemann:2000za,Wiedemann:2000tf,Wang:2001cs}.
The second feature of the parton energy loss is its dependence 
on the color representation of the propagating parton. Therefore, 
energy loss for a gluon
is 9/4 times larger than a quark. Previous works have
investigated the consequences of the second non-Abelian feature 
in the flavor dependence of the high-$p_T$ hadron 
suppression \cite{Wang:1998bh}. In this paper we study
the effect of the non-Abelian parton energy loss on
the energy dependence of the inclusive hadron spectra suppression.
We exploit the well-known feature of the initial parton
distributions in nucleons (or nuclei) that quarks dominate
at large fractional momentum ($x$) while gluons dominate
at small $x$. Jet or large $p_T$ hadron production as a result of 
hard scatterings of initial partons will be dominated by
quarks at large $x_T=2p_T/\sqrt{s}$ and by gluons at small $x_T$.
Since gluons lose 9/4 times as more energy as quarks, the
energy dependence of the large (and fixed) $p_T$ hadron spectra 
suppression due to parton energy loss should reflect the 
transition from quark-dominated jet production at low energy to
gluon-dominated jet production at high energy. Such a unique energy 
dependence of the high-$p_T$ hadron suppression can be tested by
combining $\sqrt{s}=200$ AGeV data with lower energy data
or future data from LHC experiments.

We will work within a LO pQCD parton model incorporating
the non-Abelian QCD parton energy loss in high-energy heavy-ion 
collisions. We will study the energy dependence of the high-$p_T$ 
hadron suppression and compare the effect of QCD energy loss with 
that of a non-QCD one where gluons and quarks are chosen to 
have the same amount of energy loss. In both cases, we will
assume that parton energy loss is proportional to the initial
gluon density of the system which in turn is assumed to be 
proportional to the measured total charge hadron multiplicity
in the central rapidity region.

In comparison to previous studies within the LO pQCD parton model
that employed the hard-sphere model of nuclear distribution and
assumed only longitudinal expansion, we will use more 
realistic Woods-Saxon nuclear distribution and in addition include
the tranverse expansion of the dense medium.

In a LO pQCD model \cite{Wang:1998ww}, the inclusive invariant 
differential cross section for high-$p_T$ hadrons in 
$A+B$ collisions is given by
\begin{eqnarray}
\frac{d\sigma^h_{AB}}{dyd^2p_T}
&=&K\sum_{abcd} 
\int d^2{\bf b} d^2{\bf r} dx_a dx_b 
d^2{\bf k}_{aT} d^2{\bf k}_{bT} t_A(r)t_B(|{\bf b}-{\bf r}|) 
g_A(k_{aT},r)  g_B(k_{bT},|{\bf b}-{\bf r}|) \nonumber \\
& & \times f_{a/A}(x_a,Q^2,r)f_{b/B}(x_b,Q^2,|{\bf b}-{\bf r}|)
\frac{D_{h/c}(z_c,Q^2,\Delta E_c)}{\pi z_c}  
\frac{d\sigma (ab\rightarrow cd)}{d\hat{t}}\;, 
\label{eq:nch_AA}
\end{eqnarray}
where $\sigma (ab\rightarrow cd)$ are elementary parton scattering 
cross sections. The factor $K\approx 1.0-2.0$ is used to account 
for higher order QCD corrections and is set to be the same for
both $p+p$ and $A+B$ collisions at the same energy. The hadron
is assumed to have the same rapidity as the parton,
i.e. $y=y_c$, and its fractional momentum is defined by $z_c=p_T/p_{Tc}$.
The parton distributions per nucleon $f_{a/A}(x_a,Q^2,r)$
inside the nucleus can be factorized into the parton 
distributions in a free nucleon given by the CTEQ 
parameterization \cite{Pumplin:2002vw,Stump:2003yu} 
and the impact-parameter dependent
nuclear modification factor given by the new 
HIJING parameterization:
$f_{a/A}(x_a,Q^2,r)=R^A_a(x,Q^2)[(Z/A)f_{a/p}(x,Q^2)+(1-Z/A)f_{a/n}(x,Q^2)]$ 
with $R^A_a(x,Q^2)$ given by Eqs.~(8) and (9) of Ref.\ \cite{Li:2001xa}.
We assume that the initial transverse momentum 
distribution $g_A(k_T,Q^2,b)$ has a Gaussian form 
\cite{Wang:1998ww,Dumitru:2001jx} with a width that includes both 
an intrinsic $k_T$ in a nucleon and the nuclear broadening due
to initial multiple scattering in a nucleus: 
$g_A(k_T,Q^2,b)=e^{-k_T^2/\lan k_T^2\ran _A}/(\pi \lan k_T^2\ran _A)$. 
The impact-parameter dependent broadened variance is given by 
$\lan k_T^2\ran _A(Q^2)=\lan k_T^2\ran _N (Q^2)+\delta ^2(Q^2)[\nu _A(b)-1]$, 
where the number of scatterings $\nu _A(b)$ the projectile suffers inside the 
the nucleus is $\nu _A(b)=\sigma _{NN}t_A(b)$ with the nuclear 
thickness function $t_A(b)$ defined as follows, and 
the scale-dependent $\delta ^2(Q^2)$ is chosen as 
$\delta ^2(Q^2)=0.225\ln ^2(Q/{\rm GeV})/[1+\ln (Q/{\rm GeV})]\; 
{\rm GeV}^2/c^2$. The average initial intrinsic transverse momentum 
in nucleon-nucleon collision 
is $\lan k_T^2\ran _N (Q^2)=1.2+0.2 Q^2 \alpha_s(Q^2)$. 
The scale which characterizes 
the partonic process is chosen to be $Q=p_T$, where $p_T$ 
is the transverse momentum of the final-state partons 
in a partonic scattering. Detailed description of this model and 
systematic comparisons with experimental data can be found 
in Ref.~\cite{Wang:1998ww}. In this paper we use the Woods-Saxon 
nuclear distribution $F_{WS}(r)=N_A/[1+\mathrm{exp}((r-R_A)/a)]$ 
to replace the simplified hard-sphere one as used in previous papers.
Here $R_A$ is the radius of the nucleus and given by 
$R_A=1.12A^{1.0/3.0}-0.86A^{-1.0/3.0}$, $a=0.54$ fm is 
a radius parameter and $N_A$ is the normalization constant. 
It can be further written as a function of the coordinate component 
$z$ along the beam direction of the nucleus and ${\bf b}$ that is 
perpendicular to it by $r=\sqrt{z^2+b^2}$. 
The nuclear thickness function $t_A(b)$ is then 
$t_A(b)=\int _{-\infty}^{\infty} dz F_{WS}(z,b)$ with 
the normalization condition $\int d^2b t_A(b)=A$.

The parton energy loss is encoded in an effective
modified fragmentation function \cite{Wang:1996yh,Wang:1996pe}
\begin{eqnarray}
D_{h/c}(z_c,Q^2,\Delta E_c) 
&=&(1-e^{-\langle \frac{\Delta L}{\lambda}\rangle})
\left[ \frac{z_c^\prime}{z_c} D^0_{h/c}(z_c^\prime,Q^2) 
+\langle \frac{\Delta L}{\lambda}\rangle
\frac{z_g^\prime}{z_c} D^0_{h/g}(z_g^\prime,Q^2)\right]
+ e^{-\langle\frac{\Delta L}{\lambda}\rangle} D^0_{h/c}(z_c,Q^2) \;.
\label{modfrag} 
\end{eqnarray}
This effective form is a good approximation to the actual 
calculated medium modification in the multiple parton 
scattering formalism \cite{Guo:2000nz,Wang:2001if},
given that the actual energy loss should be about 1.6 times of
the input value in the above formula. 
Here $z_c^\prime=p_T/(p_{Tc}-\Delta E_c)$,
$z_g^\prime=\langle \Delta L/\lambda\rangle p_T/\Delta E_c$
are the rescaled momentum fractions and $\Delta E_c$ is
the total energy loss during an average number 
of inelastic scatterings $\langle \Delta L/\lambda\rangle$.
The fragmentation functions in free space $D^0_{h/c}(z_c,Q^2)$
are given by the BBK parameterization \cite{Binnewies:1994ju}.

In contrast to previous calculations where only longitudinal 
expansion was considered, we incorporate in this paper both 
longitudinal and transverse expansion of the medium in the
calculation of parton energy loss. To simplify the calculation,
we use hard-sphere nuclear distribution again. 
Let us denote the gluon number $N_{g}$ and assume that $N_{g}$
is a slowly varying function of rapidity $y$ and proper time 
$\tau$ at central rapidity region $y=0$, then we have 
$d^{2}N_{g}/d\tau dy = 0$. Noting that 
$dN_{g}/dy = \rho dV/dy$ and $dV/dy=\tau \pi R_{T}^{2}$, we obtain 
\be
\frac{d\rho }{d\tau }\frac{dV}{dy}
+\rho \left[ \pi R_{T}^{2}+2\pi \tau R_{T}\frac{dR_{T}}{d\tau}\right] =0 \;,
\ee
which is
\be
\frac{d\rho }{d\tau }+\rho \left[ \frac{1}{\tau }
+\frac{2}{R_{T}}\frac{dR_{T}}{d\tau}\right] =0 \;.
\ee
The radius has the form $R_{T}(\tau )=R_{A}+(\tau -\tau _{0})c_{s}^{2}$ where
$c_{s}$ is the speed of sound in the medium 
given by $c_{s}^{2}=\partial P/\partial e $
($1/3$ for ideal gas). The solution of the above equation is then
\be
\tau \rho \left[ R_{A}+(\tau -\tau _{0})c_{s}^{2}\right] ^{2}
=\tau _{0}\rho _{0}R_{A}^{2}\;.
\ee
The expansion is characterized by the gluon density 
$\rho _g (\tau ,r)$ whose initial distribution is proportional 
to the transverse profile of participant nucleons. 
We can write the total energy loss for a parton 
traversing the medium as
\bea
\D E(b,\mathbf{r},\f )
& \approx & \lan \frac{dE}{dL}\ran _{1d}
\int _{\tau _0}^{\tau _{max}} d\tau 
\frac{\tau \left[ R_{min}+(\t -\t _0)c_s^2\right]^2-\t _0 R_{min}^2}
{\t _0R_{min}^2\rho _0} 
\rho _g (\tau ,b,\mathbf{r}+\mathbf{n}\tau ) \;,
\eea
where $R_{min}=\mathrm{Min}(R_A,R_B)$, 
$\mathbf{n}$ is the direction where a parton is propagating. 
The upper limit $\tau _{max}=\mathrm{Min}(\D L, \tau _f )$
is the longest time for the parton to propagate in the dense medium, 
where $\tau _f$ is the lifetime of the dense matter before breakup. 
$\D L (b,\mathbf{r},\f )$ is the distance 
the parton, produced at $\mathbf{r}$, travels along $\mathbf{n}$ at 
the azimuthal angle $\f$ relative to the reaction plane in a collision with 
impact parameter $b$. 
Since the formation time of a hadron fragmented 
from a parton is proportional to the energy of the parton, 
very high energetic partons generally hadronize after the 
dense medium breaks up, or they hadronize outside the medium. 
In this case we have $\tau _{max}=\D L$. 
$\lan dE/dL\ran _{1d}$ is the average parton energy loss 
over a distance $R_A$ in a 1-dimensional 
expanding medium with an initial uniform gluon density $\rho _0$.
The gluon density $\rho _g$ in the longitudinally and 
transversely expanding medium is then given by
\bea
\rho _g (\tau ,b,\mathbf{r}+\mathbf{n}\tau )
&=& \frac{\t _0\rho _0}{\t}\frac{R_{min}^2}
{\left[ R_{min}+(\t -\t _0)c_s^2\right]^2}
\frac{\p}{2c_{AB}R_{min}}\non
&&\times\left[\frac{R_A^3}{A}t_A(r)
+ \frac{R_B^3}{B}t_B(|\mathbf{b}-\mathbf{r}|)\right] .
\eea
The average number of scatterings along the path of parton 
propagation is
\bea
\lan \D L/\l \ran &=& \int _{\tau _0}^{\tau _{max}} 
d\tau \sigma \rho _g (\tau ,b,\mathbf{r}+\mathbf{n}\tau ) \;.
\eea
The energy loss function can be parameterized as
\bea
\lan \frac{dE}{dL}\ran _{1d} 
&=& \e _0 \frac{(E/\m _0-1.6)^{1.2}}{7.5+ E/\m _0} 
\eea
according to a study of parton energy loss \cite{Wang:2001cs}
that include both included bremstrahlung and thermal aborption of gluons. 
For $\sqrt{s}$=200 AGeV, we find following set of 
parameters $\e _0=1.2$ GeV, $\m_0 =1.6$ GeV and $\l _0 = 0.2$ fm
($\l _0$ appears in the formula of $\langle \Delta L/\lambda\rangle$)
can fit the data. In Ref.\ \cite{Wang:2003mm,Wang:2004yv}, 
these parameters are set to slightly different values 
$\e _0=1.07$ GeV, $\m_0 =1.5$ GeV and $\l _0 = 0.3$ fm. 
The value of $\lan \frac{dE}{dL}\ran _{1d}$ with $\e _0=1.2$ GeV, 
$\m_0 =1.6$ GeV used in this paper is almost the same 
in the energy range $E=5-20$ GeV as with previous values 
$\e _0=1.07$ GeV, $\m_0 =1.5$ GeV \cite{Wang:2003mm,Wang:2004yv}. 
For example, at $E=5$ and $20$ GeV we have
$\lan \frac{dE}{dL}\ran _{1d}(\e _0=1.2,\m_0 =1.6)=0.19$ and $1.05$, 
while $\lan \frac{dE}{dL}\ran _{1d}(\e _0=1.07,\m_0 =1.5)=0.19$ and $0.99$. 
Another modified parameter $\l _0$ in this paper 
is inversely proportional to the average number of 
scatterings undergone by the propagating
energetic parton. The smaller value $\l _0=0.2$ fm than 
previously used one $\l _0=0.3$ fm means 
that the average number of scatterings 
is tuned larger to make more energy loss by compensating the 
effect caused by transverse expansion which makes the medium 
more rapidly diluted. Note that the parameter 
$\e _0$ is proportional and $\l _0$ is inversely proportional
to the gluon or multiplicity density per rapidity. 
The energy loss in a corresponding static medium is found to 
be 14 GeV/fm, which is about 30 times as high as 
in a cold nuclei \cite{Wang:2003mm}. 

The jet quenching effect can be shown by the nuclear 
modification factor defined as \cite{Wang:2001cy}
\be
R_{AB}
=\frac{d\sigma ^h_{AB}/dyd^2p_T}{\lan N_{\rm binary}\ran
d\sigma ^h_{pp}/dyd^2p_T}\; ,
\ee
where $N_{\rm binary}$ is the average number of geometrical 
binary collisions at a given range of impact parameters
\be
\lan N_{\rm binary}\ran = 
\int d^2 {\bf b} d^2 {\bf r} t_A (r) t_B (|{\bf b}-{\bf r}|) \; .
\ee
If there is no energy loss, the cross section for nucleus-nucleus collisions 
is a simple sum of that for elementary binary nucleon-nucleon collisions, 
so the nuclear modification factor $R_{AB}$ is one. 
Hadron suppression due to parton energy loss leads to $R_{AB}<1$. 

As we mentioned earlier that we use the Woods-Saxon nuclear distribution
in the parton model calculation.
The numerical difficulty with the Woods-Saxon distribution 
is that one cannot simply put the analytical formula into the program 
becasue that would substantially slow the speed of the calculation and make 
the numerical calculation practically impossible. 
One trick to overcome this problem is to 
calculate the distribution before hand and then 
store the results in tables whose entries can be 
called in the run time of the program. 
The calculated $R_{AB}$ results for Au+Au collisions are shown 
in Figs.\ 1 and 2 with Fig.\ 1 for $\pi ^0$ and Fig.\ 2 for 
charged hadrons. The results for three collision energies 
$\sqrt{s}$=62.4, 200 and 5500 AGeV are given. 
The parameters $\epsilon _0$ and $\lambda _0$ at these energies 
are set to approriate values based on the ratios of 
charged particle (or gluon) multiplicity 
density \cite{Li:2001xa,Kharzeev:2000ph}
$(dN_{ch}/dy)_{5500}/(dN_{ch}/dy)_{200}$, 
$(dN_{ch}/dy)_{62.4}/(dN_{ch}/dy)_{200}$ and their values 
$\epsilon _0=1.2$ and $\lambda _0=0.2$ at 200 AGeV. 
In the figures one can
see different transverse momentum behaviors  
of the nuclear modification factor at these energies.
Similar behaviors have been seen in recent 
studies \cite{Wang:2004yv,Vitev:2004gn,Adil:2004cn}.
The nuclear modification factor decreases with $p_T$ 
at 62.4 AGeV, while it slightly increases with $p_T$ at 
200 AGeV. So the nuclear modification factors for neutral pions and 
charged hadrons at 62.4 AGeV 
intersect at about $p_T=11$ and 10 GeV respectively 
with those of 200 AGeV in the QCD case,  
where the energy loss parameters for gluons and quarks 
satisfy $g/q=9/4$ (we will explain this point later). 
The same feature also occurs in Ref. \cite{Wang:2004yv} 
where the hard sphere distribution 
and only the longitudinal expansion are used. 
In the intermediate $p_T$ region, 
one expects the jet fragmentation process to be
modified by other non-perturbative processes such as
parton recombination or coalescence 
\cite{Hwa:2002tu,Fries:2003vb,Greco:2003xt}. The
observed flavor dependence of the hadron suppression
and of the azimuthal anisotropy clearly points to the
effect of parton recombination that enhances both baryon and
kaon spectra in the presence of dense medium. To include this
effect in the current parton model, we have added a soft
component to kaon and baryon fragmentation function that
is proportional to the pion fragmentation function with a
weight $\sim \langle N_{\rm bin}(b,r)\rangle/[1+\exp(2p_{Tc}-15)]$ 
where $p_{Tc}$ is the transverse momemtum for parton $c$ 
(actually we have also found that the similar effect can be also 
achieved by using a function of the hadron transverse momemtum 
$p_T$: $\langle N_{\rm bin}(b,r)\rangle/[1+\exp(p_{T}-5)]$.)  
The functional form and parameters are adjusted so that
$(K+p)/\pi\approx 2$ at $p_T\sim 3$ GeV/$c$ in the most
central Au+Au collisions at $\sqrt{s}=200$ AGeV
and approaches its $p+p$ value at $p_T>5$ GeV/$c$. 
This gives rise to the splitting of the
suppression factor for charged hadrons and $\pi^0$ in the
calculation.

To study the sensitivity of hadron spectra suppression to the
non-Abelian parton energy loss, we compare the results with
two different cases at each energy: one for the QCD case where 
the energy loss for a gluon is 9/4 times as large as for a quark, 
i.e. $\Delta E_g/\Delta E_q=9/4$, the other is for a non-QCD case 
where the energy loss is chosen to be the same for both gluons and 
quarks. Similarly, the average number of inelastic 
scatterings obeys $\langle \frac{\Delta L}{\lambda}\rangle _g/
\langle \frac{\Delta L}{\lambda}\rangle _q=9/4$ in the QCD case. 
For the non-QCD case we are considering, the above ratio is set to one. 
In Figs.\ 1 and 2 one can see that 
the difference between the QCD and non-QCD cases are more 
significant for higher collision energies. This fact manifests itself 
at 200 and 5500 AGeV, where the nuclear modification factors $R_{AB}$ 
are much lower for the QCD energy loss pattern than for the non-QCD one. 
As shown in the figures, the suppression at 62.4 AGeV is 
not sensitive to gluon energy loss, but only to quark energy loss 
because of the dominance of quark jets at large $p_T$. 
At 200 AGeV, however, the suppression is sensitive both to quark
and gluon energy loss. At LHC energy, the results are 
only sensitive to gluon energy loss in the $p_T$ range we calculated. 
Such energy dependence pattern is a direct consequence of the non-Abelian 
feature of the energy loss. 

In order to demonstrate the colliding energy dependence of the nuclear 
modification factor and illustrate the difference between QCD and non-QCD
energy loss, we compute the $R_{AA}$ for neutral pions at 
fixed $p_T=6$ GeV in Au+Au collisions 
as a function of $\sqrt{s}$ from 20 AGeV to 5500 AGeV. 
Shown in Fig.\ 3 are the calculated results with both the QCD energy loss 
and a non-QCD case where the energy loss is set to be 
identical for quarks and gluons.
Two parameters $\e _0$ and $\l _0$ which are relevant to 
the energy loss are determined according to the gluon number or 
the charged particle multiplicity per rapidity
\cite{Li:2001xa,Kharzeev:2000ph}.
One can see that due to the dominant gluon bremstrahlung 
or gluon energy loss at high energy the $R_{AA}$ 
for the QCD case is more suppressed than 
the non-QCD case where the gluon energy loss is assumed to take an 
equal role as the quark one.
In the calculation, we have assumed that the lifetime of the dense
matter is equal or longer than the parton propagation time which
is essentially determined by the system size. This might not be the
case for heavy-ion collisions at lower energies, in particular
at around $\sqrt{s}=20$ AGeV. If one takes short lifetime, the
suppression factor $R_{AA}$ is much larger than
1 due to strong Cronin effect \cite{Wang:2004yv}. The dashed box 
around $\sqrt{s}=20$ AGeV in the Fig.\ 3 assumes a 
lifetime $\tau_f=0-2$ fm/$c$ and thus provides an estimate of the 
uncerntaintty due to lifetime of the dense matter. Since finite
lifetime reduce the effect of full parton energy loss, the difference
between QCD and non-QCD energy loss effect in $R_{AA}$ should be smaller.
The difference in Fig.\ 3 is therefore the upper limit.

Another interesting feature with the energy dependence of $R_{AA}$
is the change of slope around $\sqrt{s}=1300$ GeV. The rapid decrease
of $R_{AA}$ at $\sqrt{s}=20-1300$ GeV is mainly due to increase
of parton energy loss due to increased initial gluon density and also
the change of $p_T$ slope of jet production cross section with $\sqrt{s}$.
As the energy loss increases, more jets produced inside the
overlapped region are completely suppressed. Only those that
are produced within an outlayer in the overlapped region will survive.
This will be like surface emission with a finite depth.
The suppression factor $R_{AA}$ will then be determined by the width
of the outlayer which is just the averaged 
mean-free-path $\lan \lambda \ran$.
As a consequence, $R_{AA}$ will then have much weaker $\sqrt{s}$
dependence.

In summary, nuclear modification factors in 
Au+Au collisions at $\sqrt{s}$=62.4,
200 and 5500 AGeV are calculated in a LO perturbative QCD model 
with medium induced parton energy loss. 
The previous calculations based on the hard-sphere distribution 
of nucleus and the longitudinal expansion of the dense medium 
are improved in terms of a more realistic
Woods-Saxon distribution and both longitudinal and transverse expansion. 
The comparison of nuclear modification
factors for energy loss patterns in QCD and non-QCD cases shows 
sizable difference at higher colliding energies and thus
can be tested by the energy dependence of the hadron suppression
factor $R_{AA}$ in the range between $\sqrt{s}=20-1000$ GeV. 
We also found a weaker energy dependence above $\sqrt{s}=1000$ GeV
due to surface emission with finite depth.

\section*{Acknowledgements}

Q.W appreciates much help from A. Dumitru especially in the 
aspect of including the transverse expansion of the dense medium.
He also gratefully acknowledges support by 
the Center for Scientific Computing (CSC) 
at University of Frankfurt where most of the numerical 
calculation was done. X.-N.W. thanks J.C. Peng for helpful discussions
that stimulated the idea of testing the non-Abelian parton energy loss.
This work was supported by the Director, 
Office of Energy Research, Office of High Energy and Nuclear Physics, 
Divisions of Nuclear Physics, of the U.S. Department of Energy under 
Contract No.\ DE-AC03-76SF00098

\begin{figure}

\includegraphics[scale=0.5,angle=-90]{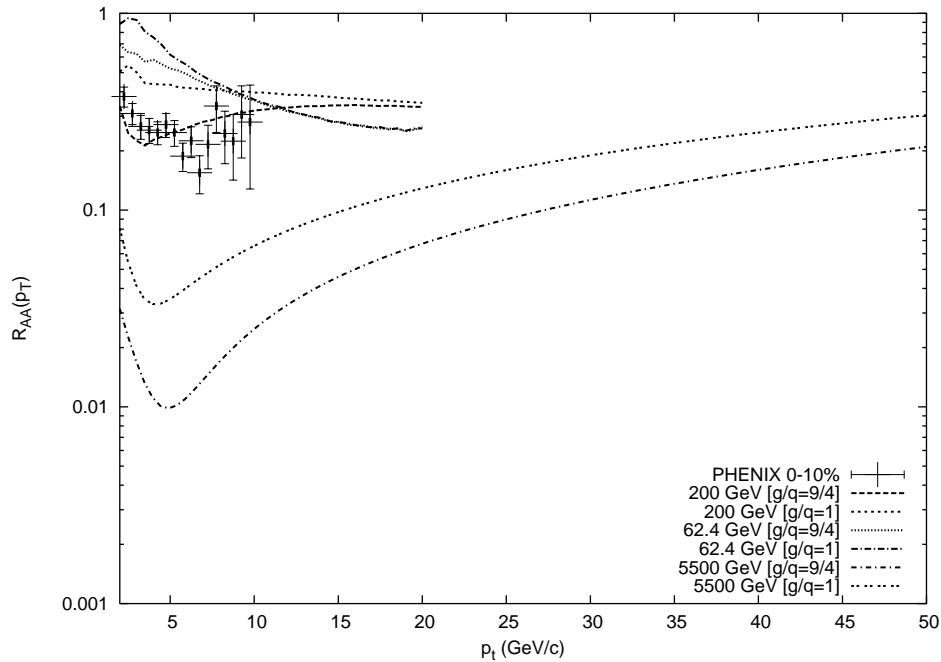}
\caption{Nuclear modification factor $R_{AuAu}$ for neutral pions at 
$\sqrt{s}=$62.4, 200 and 5500 AGeV. We choose the values of corresponding 
parameters at 62.4 and 5500 AGeV based on their values  
at 200 AGeV and the ratio 
$(dN_{ch}/dy)_{62}/(dN_{ch}/dy)_{200}$ and 
$(dN_{ch}/dy)_{5500}/(dN_{ch}/dy)_{200}$. }

\end{figure}

\begin{figure}

\includegraphics[scale=0.5,angle=-90]{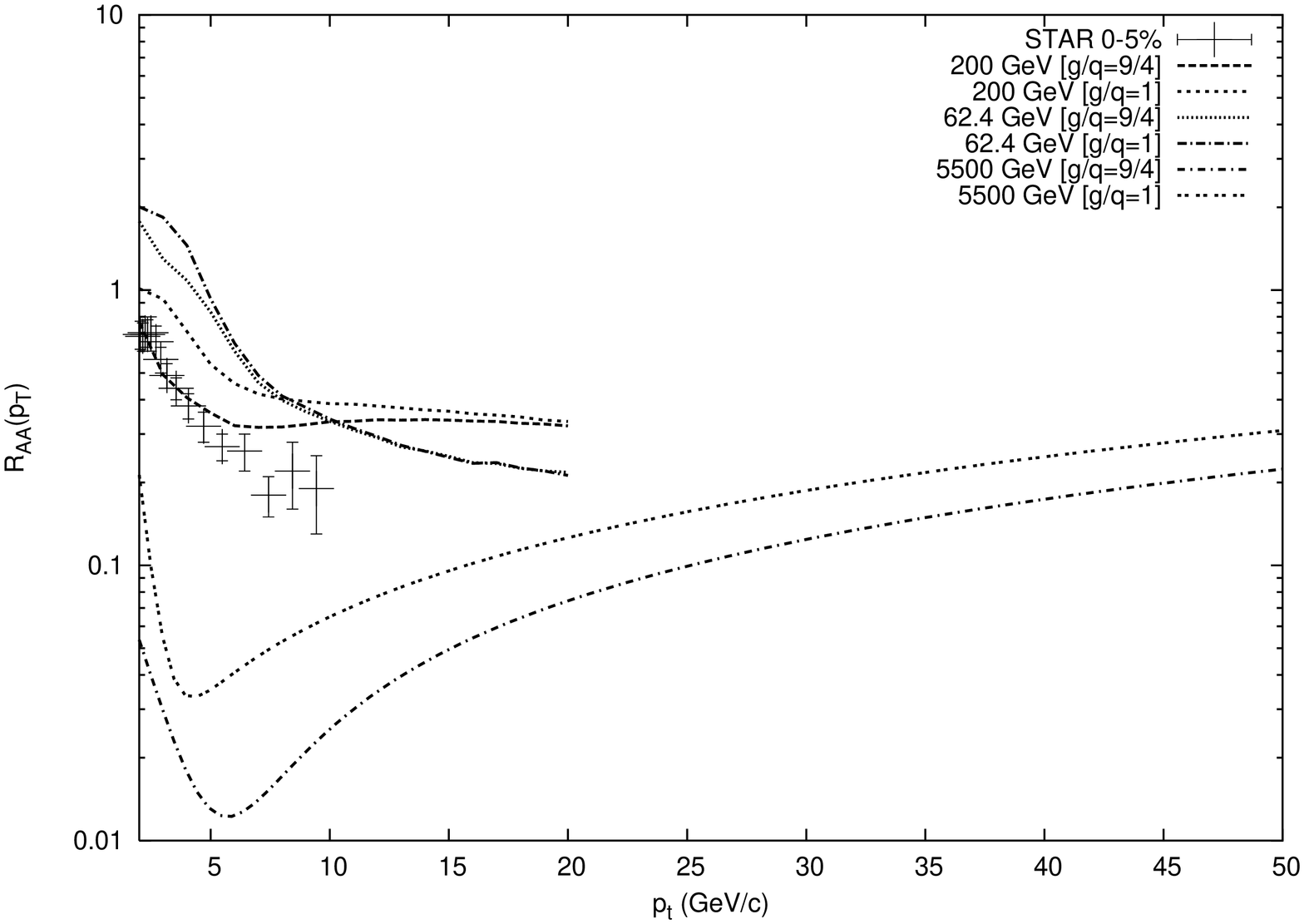}
\caption{Nuclear modification factor $R_{AuAu}$ for hadrons at
$\sqrt{s}=$62.4, 200 and 5500 AGeV. 
We choose the values of corresponding 
parameters at 62.4 and 5500 AGeV based on their values at 200 AGeV 
and the ratio $(dN_{ch}/dy)_{62}/(dN_{ch}/dy)_{200}$ and 
$(dN_{ch}/dy)_{5500}/(dN_{ch}/dy)_{200}$. }

\end{figure}

\begin{figure}

\includegraphics[scale=0.5,angle=-90]{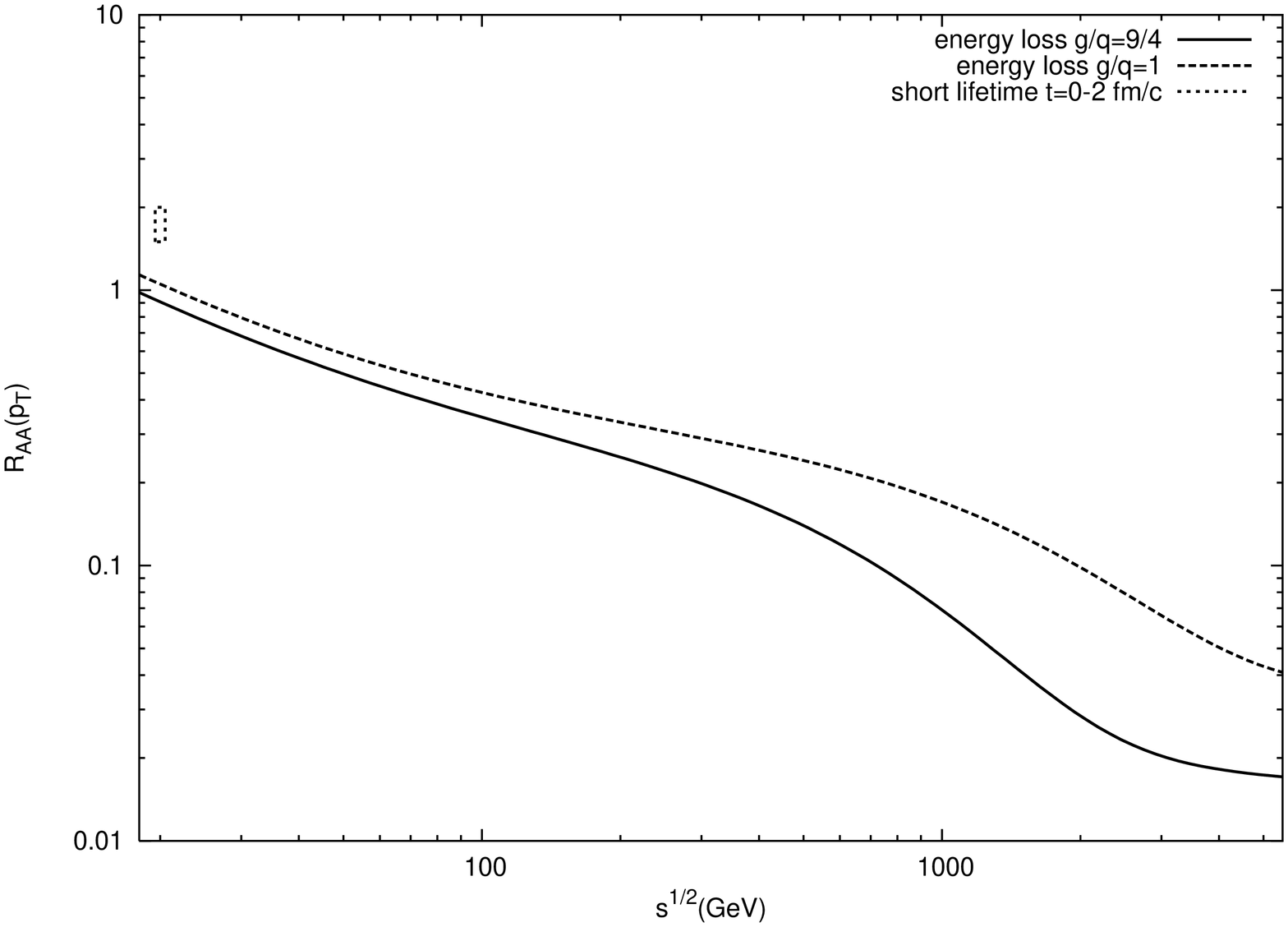}
\caption{Nuclear modification factor $R_{AuAu}$ for neutral pions 
as function of collision energy at fixed $p_T=6$ GeV in most central 
collisions (with centrality 10\%). Here we compare the QCD energy loss 
and a non-QCD one where the energy loss is identical for quarks and gluons.}

\end{figure}


\begin{thebibliography}{1}

\bibitem{Adcox:2001jp}
K.~Adcox {\it et al.}  [PHENIX Collaboration],
Phys.\ Rev.\ Lett.\  {\bf 88}, 022301 (2002)
[arXiv:nucl-ex/0109003].

\bibitem{Adler:2002xw}
C.~Adler {\it et al.},
Phys.\ Rev.\ Lett.\  {\bf 89}, 202301 (2002)
[arXiv:nucl-ex/0206011].

\bibitem{Adler:2002tq}
C.~Adler {\it et al.}  [STAR Collaboration],
Phys.\ Rev.\ Lett.\  {\bf 90}, 082302 (2003)
[arXiv:nucl-ex/0210033].

\bibitem{Adler:2002ct}
C.~Adler {\it et al.}  [STAR Collaboration],
Phys.\ Rev.\ Lett.\  {\bf 90}, 032301 (2003)
[arXiv:nucl-ex/0206006].


\bibitem{Adler:2001bp}
C.~Adler {\it et al.}  [the STAR Collaboration],
Phys.\ Rev.\ Lett.\  {\bf 86}, 4778 (2001)
[Erratum-ibid.\  {\bf 90}, 119903 (2003)]
[arXiv:nucl-ex/0104022].



\bibitem{Adcox:2001mf}
K.~Adcox {\it et al.}  [PHENIX Collaboration],
Phys.\ Rev.\ Lett.\  {\bf 88}, 242301 (2002)
[arXiv:nucl-ex/0112006].

\bibitem{Adler:2002pba}
C.~Adler {\it et al.}  [STAR Collaboration],
Phys.\ Rev.\ Lett.\  {\bf 89}, 092301 (2002)
[arXiv:nucl-ex/0203016].

\bibitem{Adcox:2002au}
K.~Adcox {\it et al.}  [PHENIX Collaboration],
Phys.\ Rev.\ Lett.\  {\bf 89}, 092302 (2002)
[arXiv:nucl-ex/0204007].



\bibitem{Hwa:2002tu}
R.~C.~Hwa and C.~B.~Yang,
Phys.\ Rev.\ C {\bf 67}, 034902 (2003)
[arXiv:nucl-th/0211010].


\bibitem{Fries:2003vb}
R.~J.~Fries, B.~Muller, C.~Nonaka and S.~A.~Bass,
Phys.\ Rev.\ Lett.\  {\bf 90}, 202303 (2003)
[arXiv:nucl-th/0301087].

\bibitem{Greco:2003xt}
V.~Greco, C.~M.~Ko and P.~Levai,
Phys.\ Rev.\ Lett.\  {\bf 90}, 202302 (2003)
[arXiv:nucl-th/0301093].

\bibitem{Voloshin:2002wa}
S.~A.~Voloshin,
Nucl.\ Phys.\ A {\bf 715}, 379 (2003)
[arXiv:nucl-ex/0210014].







\bibitem{Adams:2003zg}
J.~Adams {\it et al.}  [STAR Collaboration],
Phys.\ Rev.\ Lett.\  {\bf 92}, 062301 (2004)
[arXiv:nucl-ex/0310029].

\bibitem{Adams:2003am}
J.~Adams {\it et al.}  [STAR Collaboration],
Phys.\ Rev.\ Lett.\  {\bf 92}, 052302 (2004)
[arXiv:nucl-ex/0306007].

\bibitem{Adler:2002pu}
C.~Adler {\it et al.}  [STAR Collaboration],
Phys.\ Rev.\ C {\bf 66}, 034904 (2002)
[arXiv:nucl-ex/0206001].

\bibitem{Adler:2003kt}
S.~S.~Adler {\it et al.}  [PHENIX Collaboration],
Phys.\ Rev.\ Lett.\  {\bf 91}, 182301 (2003)
[arXiv:nucl-ex/0305013].

\bibitem{Stocker:1979bi}
H.~Stocker, J.~A.~Maruhn and W.~Greiner,
Z.\ Phys.\ A {\bf 293}, 173 (1979).

\bibitem{Csernai:1981nq}
L.~Csernai and H.~Stocker,
Phys.\ Rev.\ C {\bf 25}, 3208 (1981).

\bibitem{Ollitrault:1992bk}
J.~Y.~Ollitrault,
Phys.\ Rev.\ D {\bf 46}, 229 (1992).

\bibitem{Molnar:2001ux}
D.~Molnar and M.~Gyulassy,
Nucl.\ Phys.\ A {\bf 697}, 495 (2002)
[Erratum-ibid.\ A {\bf 703}, 893 (2002)]
[arXiv:nucl-th/0104073].

\bibitem{Zhang:1999rs}
B.~Zhang, M.~Gyulassy and C.~M.~Ko,
Phys.\ Lett.\ B {\bf 455}, 45 (1999)
[arXiv:nucl-th/9902016].

\bibitem{Xu:2004mz}
Z.~Xu and C.~Greiner,
arXiv:hep-ph/0406278.






\bibitem{Gyulassy:2004zy}
M.~Gyulassy and L.~McLerran,
arXiv:nucl-th/0405013.

\bibitem{Jacobs:2004qv}
P.~Jacobs and X.~N.~Wang,
arXiv:hep-ph/0405125.


\bibitem{Adams:2003im}
J.~Adams {\it et al.}  [STAR Collaboration],
Phys.\ Rev.\ Lett.\  {\bf 91}, 072304 (2003).

\bibitem{Adler:2003ii}
S.~S.~Adler {\it et al.}  [PHENIX Collaboration],
Phys.\ Rev.\ Lett.\  {\bf 91}, 072303 (2003).

\bibitem{Back:2003ns}
B.~B.~Back {\it et al.}  [PHOBOS Collaboration],
Phys.\ Rev.\ Lett.\  {\bf 91}, 072302 (2003).

\bibitem{Arsene:2003yk}
I.~Arsene {\it et al.}  [BRAHMS Collaboration],
Phys.\ Rev.\ Lett.\  {\bf 91}, 072305 (2003)
[arXiv:nucl-ex/0307003].




\bibitem{Wang:2003aw}
X.~N.~Wang,
Phys.\ Lett.\ B {\bf 579}, 299 (2004)
[arXiv:nucl-th/0307036].

\bibitem{Wang:xy}
X.~N.~Wang and M.~Gyulassy,
Phys.\ Rev.\ Lett.\  {\bf 68}, 1480 (1992).

\bibitem{Pluemer:dp}
M.~Pluemer, M.~Gyulassy and X.~N.~Wang,
Nucl.\ Phys.\ A {\bf 590}, 511C (1995).

\bibitem{Wang:2000fq}
X.~N.~Wang,
Phys.\ Rev.\ C {\bf 63}, 054902 (2001)
[arXiv:nucl-th/0009019].

\bibitem{Gyulassy:2000gk}
M.~Gyulassy, I.~Vitev and X.~N.~Wang,
Phys.\ Rev.\ Lett.\  {\bf 86}, 2537 (2001)
[arXiv:nucl-th/0012092].

\bibitem{Eskola:2004cr}
K.~J.~Eskola, H.~Honkanen, C.~A.~Salgado and U.~A.~Wiedemann,
arXiv:hep-ph/0406319.


\bibitem{Wang:2003mm}
X.~N.~Wang,
Phys.\ Lett.\ B {\bf 595}, 165 (2004)
[arXiv:nucl-th/0305010].



\bibitem{Vitev:2004bh}
I.~Vitev,
arXiv:hep-ph/0403089.










\bibitem{Wang:1998ww}
X.~N.~Wang,
Phys.\ Rev.\ C {\bf 61}, 064910 (2000)
[arXiv:nucl-th/9812021].


\bibitem{Wang:1998bh}
X.~N.~Wang,
Phys.\ Rev.\ C {\bf 58}, 2321 (1998)
[arXiv:hep-ph/9804357].

\bibitem{Wang:2004yv}
X.~N.~Wang,
Phys.\ Rev.\ C {\bf 70}, 031901 (2004)
[arXiv:nucl-th/0405029].

\bibitem{Vitev:2002pf}
I.~Vitev and M.~Gyulassy,
Phys.\ Rev.\ Lett.\  {\bf 89}, 252301 (2002)
[arXiv:hep-ph/0209161].

\bibitem{Gyulassy:1993hr}
M.~Gyulassy and X.~n.~Wang,
Nucl.\ Phys.\ B {\bf 420}, 583 (1994)
[arXiv:nucl-th/9306003].

\bibitem{Wang:1994fx}
X.~N.~Wang, M.~Gyulassy and M.~Plumer,
Phys.\ Rev.\ D {\bf 51}, 3436 (1995)
[arXiv:hep-ph/9408344].



\bibitem{Gyulassy:2000fs}
M.~Gyulassy, P.~Levai and I.~Vitev,
Phys.\ Rev.\ Lett.\  {\bf 85}, 5535 (2000)
[arXiv:nucl-th/0005032].

\bibitem{Gyulassy:1999zd}
M.~Gyulassy, P.~Levai and I.~Vitev,
Nucl.\ Phys.\ B {\bf 571}, 197 (2000)
[arXiv:hep-ph/9907461].

\bibitem{Baier:2001yt}
R.~Baier, Y.~L.~Dokshitzer, A.~H.~Mueller and D.~Schiff,
JHEP {\bf 0109}, 033 (2001)
[arXiv:hep-ph/0106347].

\bibitem{Wiedemann:2000za}
U.~A.~Wiedemann,
Nucl.\ Phys.\ B {\bf 588}, 303 (2000)
[arXiv:hep-ph/0005129].

\bibitem{Wiedemann:2000tf}
U.~A.~Wiedemann,
Nucl.\ Phys.\ A {\bf 690}, 731 (2001)
[arXiv:hep-ph/0008241].

\bibitem{Wang:2001cs}
E.~Wang and X.~N.~Wang,
Phys.\ Rev.\ Lett.\  {\bf 87}, 142301 (2001)
[arXiv:nucl-th/0106043].






\bibitem{Pumplin:2002vw}
J.~Pumplin, D.~R.~Stump, J.~Huston, H.~L.~Lai, P.~Nadolsky and W.~K.~Tung,
JHEP {\bf 0207}, 012 (2002)
[arXiv:hep-ph/0201195].

\bibitem{Stump:2003yu}
D.~Stump, J.~Huston, J.~Pumplin, W.~K.~Tung, H.~L.~Lai, 
S.~Kuhlmann and J.~F.~Owens,
JHEP {\bf 0310}, 046 (2003)
[arXiv:hep-ph/0303013].

\bibitem{Li:2001xa}
S.~Y.~Li and X.~N.~Wang,
Phys.\ Lett.\ B {\bf 527}, 85 (2002)
[arXiv:nucl-th/0110075].

\bibitem{Dumitru:2001jx}
A.~Dumitru, L.~Frankfurt, L.~Gerland, H.~Stocker and M.~Strikman,
Phys.\ Rev.\ C {\bf 64}, 054909 (2001)
[arXiv:hep-ph/0103203].



\bibitem{Wang:1996yh}
X.~N.~Wang, Z.~Huang and I.~Sarcevic,
Phys.\ Rev.\ Lett.\  {\bf 77}, 231 (1996)
[arXiv:hep-ph/9605213].

\bibitem{Wang:1996pe}
X.~N.~Wang and Z.~Huang,
Phys.\ Rev.\ C {\bf 55}, 3047 (1997)
[arXiv:hep-ph/9701227].

\bibitem{Guo:2000nz}
X.~F.~Guo and X.~N.~Wang,
Phys.\ Rev.\ Lett.\  {\bf 85}, 3591 (2000)
[arXiv:hep-ph/0005044].

\bibitem{Wang:2001if}
X.~N.~Wang and X.~F.~Guo,
Nucl.\ Phys.\ A {\bf 696}, 788 (2001)
[arXiv:hep-ph/0102230].

\bibitem{Binnewies:1994ju}
J.~Binnewies, B.~A.~Kniehl and G.~Kramer,
Z.\ Phys.\ C {\bf 65}, 471 (1995)
[arXiv:hep-ph/9407347].

\bibitem{Wang:2001cy}
E.~Wang and X.~N.~Wang,
Phys.\ Rev.\ C {\bf 64}, 034901 (2001)
[arXiv:nucl-th/0104031].

\bibitem{Vitev:2004gn}
I.~Vitev,
arXiv:nucl-th/0404052.

\bibitem{Adil:2004cn}
A.~Adil and M.~Gyulassy,
arXiv:nucl-th/0405036.

\bibitem{Kharzeev:2000ph}
D.~Kharzeev and M.~Nardi,
Phys.\ Lett.\ B {\bf 507}, 121 (2001)
[arXiv:nucl-th/0012025].

\end{thebibliography}
\end{document}